\documentclass[prl,letterpaper,twocolumn,preprintnumbers,nofootinbib]{revtex4-1}

\usepackage{slashed}
\usepackage{amsmath,amssymb}
\usepackage{bbm}
\usepackage{graphicx}
\usepackage{units}
\usepackage{bbold}
\usepackage{xcolor}
\usepackage{dsfont}
\definecolor{naviBlue}{RGB}{0,0,128}
\usepackage[colorlinks  = true,
            linkcolor   = naviBlue,
            urlcolor    = naviBlue,
            citecolor   = naviBlue,
            anchorcolor = naviBlue]{hyperref}

\usepackage{comment}

\newcommand{\beq}{\begin{equation}}
\newcommand{\eeq}{\end{equation}}
\newcommand{\bea}{\begin{eqnarray}}
\newcommand{\ena}{\end{eqnarray}}
\newcommand{\dd}{{\rm d}}

\newcommand{\avb}[1]{\big\langle #1 \big\rangle}	
\newcommand{\eq}{\mathrm{eq}}
\newcommand{\ie}{\emph{i.e.}}
\newcommand{\eg}{\emph{e.g.}}

\begin{document}

\title{Conversion-Driven Leptogenesis:
A Testable Theory of Dark Matter \\
and Baryogenesis at the Electroweak Scale
}

\author{Jan Heisig}
\email[E-mail: ]{heisig@physik.rwth-aachen.de}
\thanks{\\ ORCID: \href{https://orcid.org/0000-0002-7824-0384}{0000-0002-7824-0384}.}
\affiliation{\footnotesize Institute for Theoretical Particle Physics and Cosmology, RWTH Aachen University, 52056 Aachen, Germany\\ \looseness=-1 and Department of Physics, University of Virginia, Charlottesville, Virginia 22904-4714, USA \looseness=-1}

\begin{abstract}
The phenomena of dark matter and the baryon asymmetry pose two of the most pressing questions in  today's fundamental physics. Conversion-driven freeze-out has emerged as a successful mechanism to generate the observed dark matter relic density. It supports  thermalization of dark matter despite its very weak couplings, aligning with the null results from direct and indirect detection experiments. In this Letter, we demonstrate that the departure from equilibrium of dark matter, induced by semi-efficient conversions, satisfies Sakharov’s conditions, providing a novel explanation for both dark matter and the baryon asymmetry. Specifically, for a leptophilic model, we establish the mechanism of conversion-driven leptogenesis. The scenario predicts dark matter masses close to the electroweak scale offering testable predictions, such as soft displaced leptons at the LHC and future colliders.
\end{abstract}

\maketitle

\emph{\bfseries Introduction}---The puzzles posed by dark matter (DM) and the baryon asymmetry in our Universe (BAU) remain among the most compelling challenges in modern physics. Each problem hints at physics beyond the standard model (SM) and has sparked extensive investigations both theoretically and experimentally~\cite{Bertone:2004pz,Davidson:2008bu}.
While numerous mechanisms have been proposed to tackle these puzzles individually, there is an appealing notion of linking the two phenomena together, notably explored within the frameworks of 
sterile neutrino models~\cite{Asaka:2005pn,Canetti:2012vf} or asymmetric DM~\cite{Petraki:2013wwa}.
In this Letter, we present a complementary approach to simultaneously addressing these phenomena, economically yielding DM thermalization, lepton asymmetry generation, and DM freeze-out from the same processes, without ever generating a DM asymmetry.
Specifically, we demonstrate that conversion-driven freeze-out (or coscattering)~\cite{Garny:2017rxs,DAgnolo:2017dbv} is capable of generating the BAU in a mechanism we call \emph{conversion-driven leptogenesis}. 

Conversion-driven freeze-out has proven to be effective in explaining the observed abundance of DM. Due to the small DM couplings required in this scenario, DM is thermally coupled through conversion processes only. Hence, freeze-out occurs through the cessation of efficient conversions rather than annihilation processes. Interestingly, due to their specific scaling with temperature, conversion processes are semi-efficient already when DM is semi-relativistic leading to a deviation from thermal equilibrium before the onset of significant DM dilution. 
This early out-of-equilibrium phase is key for fulfilling Sakharov's conditions~\cite{Sakharov:1967dj} in conversion-driven leptogenesis. 
It opens up the intriguing possibility of generating a sufficiently large lepton asymmetry provided the existence of $CP$-violating couplings entering the conversion processes.
Sphalerons~\cite{Kuzmin:1985mm} then convert this lepton asymmetry into a baryon asymmetry.

The simplest realizations of conversion-driven freeze-out require two new particles, a DM candidate and a coannihilating partner such as a `$t$-channel mediator'.
Yet, to produce a non-zero asymmetry in the coannihilator's decay into DM, an additional physical state is necessary.
While several possibilities exist, we consider a DM multiplet of a (global) symmetry as considered in lepton-flavored DM~\cite{Chen:2015jkt,Acaroglu:2022hrm}. 
This framework serves as a minimal representation of a more complete theory, potentially offering links to other outstanding problems of the SM such as the origin of neutrino masses and the flavor puzzle.
For instance, an embedding in radiative neutrino mass models~\cite{Ma:2006km}\footnote{See, \eg, Refs.~\cite{Hugle:2018qbw,Seto:2022tow} for links to the BAU in these models.} appears promising, 
since conversion-driven freeze-out 
unveils an interesting region of its parameter space~\cite{Heeck:2022rep}.


\vspace{1.6ex}
\emph{\bfseries Model and Boltzmann equations}---We consider a simplified leptophilic $t$-channel mediator DM model introducing the scalar mediator $\phi$ with the same gauge quantum numbers as the right-handed charged leptons and a Majorana multiplet $\chi$ with mass eigenstates $\chi_i$, the lightest of which, $\chi_1$, constitutes the DM candidate. These new particles are assumed to be odd under a new $Z_2$ symmetry while SM particles are even. The  
Lagrangian reads:
\begin{align}
    \mathcal{L} \,\supset\, & \frac{1}{2}  \bar{\chi_i}  \,\slash\!\!\!\!\!\;\partial \,\chi_i-\frac{M_\chi^{ij}}{2} \bar{\chi_i} \chi_j+ (D_{\mu}\phi)^\dagger \, D^{\mu}\phi -m_\phi^2 |\phi|^2 \nonumber \\
    &- \
  (\lambda_{ki} \phi  \,\bar \ell_{\text{R},k} \chi_i  \ + \ \text{h.c.} ) -\lambda_H H^{\dag} H \phi^{\dag} \phi\,,
\end{align}
where $m_\phi$ is the scalar mass, $M_\chi=\text{diag}(m_{\chi_1},\dots, m_{\chi_n})$ is the diagonal Majorana mass matrix, $\lambda$ a complex Yukawa matrix, $\lambda_H$ the Higgs-portal coupling, and 
$D_\mu$ the covariant derivative.
In the context of conversion-driven freeze-out, Ref.~\cite{Junius:2019dci} studied a similar model but with one Majorana fermion only; a realization for flavored (but quarkphilic) DM has been studied in \cite{Acaroglu:2023phy}.

In the regime of conversion-driven freeze-out, $\phi$ pair-annihilation mediated by its gauge interactions is the only process supporting the dilution of the $Z_2$-odd sector's abundance in the early Universe. Conversions between $\phi$ and $\chi_i$, in turn, govern the abundance of the $\chi_i$. The leading conversion processes are the decays and inverse decays $\phi\leftrightarrow \chi_i \ell_\text{R}$.
In the presence of complex phases in $\lambda_{ik}$, this decay gives rise to $CP$-violation. We introduce the $CP$-asymmetry
\begin{equation}
    \label{eq:epsil}
    \epsilon_i  \equiv \frac{\Gamma_i^- -\Gamma_i^+}{2 \Gamma_{i}} \,,
\end{equation}
where $\Gamma^\pm_i=\sum_k\Gamma (\phi^\pm\!\to {\chi_i} \ell_k^\pm)$ and $\Gamma_i = (\Gamma^-_i\! + \Gamma^+_i)/2$. (Below we  also consider $\Gamma_{\!ik}$ which is the same quantity as $\Gamma_i$ but  without the sum over lepton flavors $k$, as well as $\Gamma_\text{tot} = \sum_i \Gamma_i$.)
Note that the $\epsilon_i$ are not fully independent of each other as $CPT$ invariance requires the total decay widths of $\phi^+$ and $\phi^-$ to be equal.

Further introducing $\Sigma_A=Y_A+Y_{\bar{A}}$ and $\Delta_A=Y_A-Y_{\bar{A}}$, we can formulate the Boltzmann equations for the relevant abundances and asymmetries as
\begin{align}
    \label{eq:BMESigPhi}
  \frac{\dd \Sigma_{\phi}}{\dd x} &= \frac{1}{ 3 \mathcal{H}}\frac{\dd s}{\dd x}
  \left[\,\frac12 \avb{\sigma v}_{\phi^+\! \phi^-}\left(\Sigma_{\phi}^2- {\Sigma_{\phi}^{\eq}}^{2}\right)  \right.\\ \nonumber
 &\quad \left.+\sum_i\frac{\Gamma_i }{s}\left(\Sigma_{\phi}-\Sigma_{\phi}^\eq \frac{Y_{\chi_i}}{Y_{\chi_i}^{\eq}}\right)\right], \\
\label{eq:BMESigChi}
  \frac{\dd Y_{\chi_i}}{\dd x} &=  -\frac{1}{ 3 s \mathcal{H}}\frac{\dd s}{\dd x}
  \Gamma_i\left(\Sigma_{\phi}-\Sigma_{\phi}^\eq \frac{Y_{\chi_i} }{Y_{\chi_i}^{\eq} }\right) ,
  \\
     \label{eq:BMEDelPhi}
  \frac{\dd \Delta_{\phi}}{\dd x} &= \frac{1}{ 3 s \mathcal{H}}\frac{\dd s}{\dd x}\sum_i
  \left[
  \Gamma_i\epsilon_i \left(\Sigma_{\phi}-\Sigma_{\phi}^\eq \frac{Y_{\chi_i} }{Y_{\chi_i}^{\eq} }\right)
   \right.\\ \nonumber
 &\quad \left.
  +\,\Gamma_i\Delta_{\phi}- \sum_k \Gamma_{ik}\Delta_{\ell_k} \frac{Y_{\chi_i} \Sigma_{\phi}^\eq}{Y_{\chi_i}^\eq \Sigma_{\ell_k}^\eq}
  \right],
\end{align}
as well as an equation for each $\dd \Delta_{\ell_k}/\dd x$ whose left-hand side is equal to the one of Eq.~\eqref{eq:BMEDelPhi} but with opposite sign and without the sum over $k$ (implicit for $\Gamma_i,\epsilon_i$) such that $\sum_k\dd \Delta_{\ell_k}/\dd x = -\dd \Delta_{\phi}/\dd x$. 
Here, $x=m_{\chi_1}/T$, $\mathcal{H}$ is the Hubble expansion rate,  and $s$ is entropy density. We have neglected terms of order $\epsilon_i^2$ and $\epsilon_i \Delta_A$. 
Note that $\Sigma_{\ell}=\Sigma_{\ell}^\eq$ due to efficient gauge interactions with the SM bath and that $\chi_i$ is its own antiparticle, \ie~$\Delta_{\chi_i}=0$.


\vspace{1.6ex}
\emph{\bfseries Cosmologically viable realizations}---To provide a proof-of-principle of the cosmological viability of the mechanism, we consider two $\chi_i$ assuming quasi-mass degeneracy, $\Delta m_{12}\lll m_{\chi}$, where $m_\chi$ is the common mass and $\Delta m_{12}$ the mass splitting. 
For concreteness, we choose couplings to the first two generations of SM leptons only, $\ell_{\text{R},k}=e_\text{R},\mu_\text{R}$ such that $\lambda$ is a $2\times 2$ matrix. 
In this case, $CPT$ invariance yields $\epsilon_2 \mathcal{B}_2 = - \epsilon_1 \mathcal{B}_1$, where $\mathcal{B}_i = \Gamma_i/\Gamma_\text{tot}$ 
and
$\mathcal{B}_2=1-\mathcal{B}_1$. By restricting ourselves to 
the case $\Gamma_{i1}\sim \Gamma_{i2}$, we can further approximate the last term in Eq.~\eqref{eq:BMEDelPhi} in the limit of massless SM leptons by $-\Gamma_i \Delta_\ell Y_{\chi_i} \Sigma_\phi^\eq/ (Y_{\chi_i}^\eq \Sigma_\ell^\eq)$, where $\Delta_\ell=\sum_k \Delta_{\ell_k}$,  $\Sigma_\ell^\eq =\sum_k \Sigma_{\ell_k}^\eq$. Upon vanishing initial asymmetries, $\Delta_\ell=-\Delta_\phi$, and we only need to solve the displayed set of (four) Boltzmann equations \eqref{eq:BMESigPhi}--\eqref{eq:BMEDelPhi}. 
Dividing Eq.~\eqref{eq:BMEDelPhi} by $\epsilon \equiv \epsilon_1$ allows us to factor out its $\epsilon$-dependence. In this setup, we encounter a total of five free model parameters, $m_\phi$, $m_\chi$, $\lambda_H$, $\Gamma_\text{tot}$, and $\mathcal{B}_1$, entering the numerical solutions for $\Sigma_\phi$, $Y_{\chi_i}$, and $\Delta_\phi/\epsilon$.

We compute the annihilation and conversion processes with MadDM~\cite{Ambrogi:2018jqj} and MadGraph5\_aMC@NLO~\cite{Alwall:2014hca} and take into account Sommerfeld enhancement of the $s$-wave contribution of the annihilation cross section. 
We also include effects of (excited) bound states following Refs.~\cite{Garny:2021qsr,Binder:2023ckj} which, however, turned out to have a minor
effect on our results. 
While, according to Eq.~\eqref{eq:epsil}, we only consider the asymmetry from the decays, we also include the (symmetric contribution from) $2\to2$ scattering to the conversions in our numerical solution taking into account the leading processes $\phi \ell \to \chi_k \gamma$, $\phi \gamma \to \chi_k \ell$. The latter process is regulated by a thermal photon mass. A more consistent treatment of the thermal corrections has recently been presented in Ref.~\cite{Becker:2023vwd}. While their effect on the freeze-out abundance is expected to be small, they may enhance the asymmetry.

\begin{figure}
  \centering
\vspace{2ex}
\includegraphics[width=0.45\textwidth, trim= {0.0cm 0.0cm 0.0cm 0.12cm}, clip]{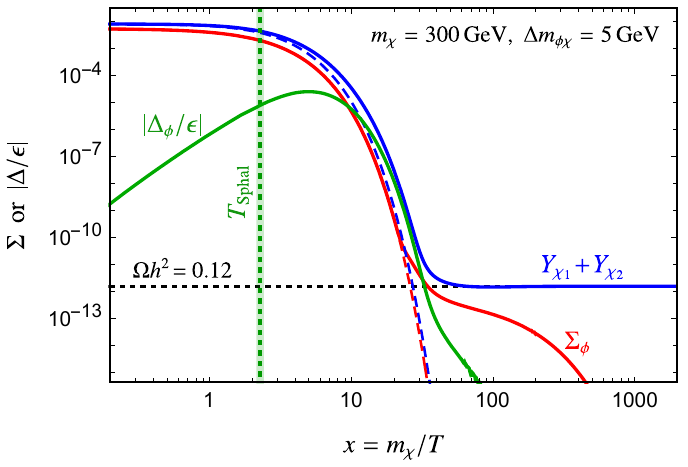} 
\vspace{-1ex}
  \caption{
Solutions of the Boltzmann equations for a cosmologically valid parameter point, $\Omega h^2 = 0.12$, with $\mathcal{B}_1=0.95$ and $\lambda_H=1$.  The dashed curves denote the corresponding equilibrium abundances.
The vertical green, dotted line denotes the sphaleron decoupling temperature $T_\text{Sphal}\simeq132$\,GeV. 
}
  \label{fig:BMEsol}
\end{figure}

We find cosmologically viable scenarios that explain the measured relic density, $\Omega h^2=0.12$~\cite{Planck:2018vyg}, in a wide range of parameter space, ranging up to DM masses of around 510\,(640)\,GeV for a mass splitting $\Delta m_{\phi\chi}\equiv m_\phi-m_\chi$ of 5\,(2.5)\,GeV and for $\lambda_H=1$. (For smaller (larger) values of $\lambda_H$, the allowed parameter space shrinks (widens) due to the smaller (larger) $\phi$-pair annihilation rate. For $\lambda_H=0$, annihilation proceeds via the $\phi$'s gauge interactions only, restricting the solutions to $m_\chi < 150\,$GeV.) The required entries of the coupling matrix $\lambda$ are of the order of $10^{-6}$, similar to what has been found in Ref.~\cite{Garny:2017rxs}.
The evolution of the number densities and the asymmetry $|\Delta_\phi/\epsilon|$ are shown in Fig.~\ref{fig:BMEsol} for an exemplary point in 
parameter space.
The asymmetry (green curve) is built up due to a gradually increasing departure from equilibrium of $\chi_i$. It reaches a maximum around $x \sim 5$ after which it decreases again due to the decrease in the number density of $\phi$. 
This result is largely insensitive to the initial conditions at $x\ll 1$. Interestingly, for a DM mass of a few hundred GeV, the point of sphaleron decoupling, $T_\text{Sphal}\simeq 132\,$GeV~\cite{DOnofrio:2014rug}, at which the baryon-asymmetry is frozen, coincides favorably with the maximum in $\Delta_\phi$. This is a crucial observation and leads to the prediction of sizable values of $|\Delta_\phi/\epsilon|_{T=T_\text{Sphal}}$.


\vspace{1.6ex}
\emph{\bfseries Baryon asymmetry}---Following the lepton asymmetry generation
(and left-right equilibration), efficient sphaleron processes erase $B+L$, converting $\Delta_\ell$ into a baryon asymmetry~\cite{Kuzmin:1985mm,Harvey:1990qw}:
\begin{align}
    Y_{\Delta B} \simeq \frac{28}{79} \Delta_\ell = -\frac{28}{79}  \epsilon \times \left.\frac{\Delta_\phi}{\epsilon}\right|_{T=T_\text{Sphal}}.
    \label{eq:YDeltaB}
\end{align}
For the considered slices in parameter space, $|\Delta_\phi/\epsilon|_{T_\text{Sphal}}$ lies between  $10^{-6}$ and $10^{-5}$ for DM mass up to 450\,GeV (and falls off towards larger masses). To explain the observed BAU, $Y_{\Delta B}^\text{meas} \simeq 0.9\times 10^{-10}$~\cite{Davidson:2008bu,Planck:2018vyg}, this requires $|\epsilon|\sim 10^{-5}$--$10^{-4}$ (or larger).

\begin{figure}[tb]
    \centering
    \includegraphics[scale=0.18, trim= {7.5cm 6.5cm 5cm 5cm}, clip]{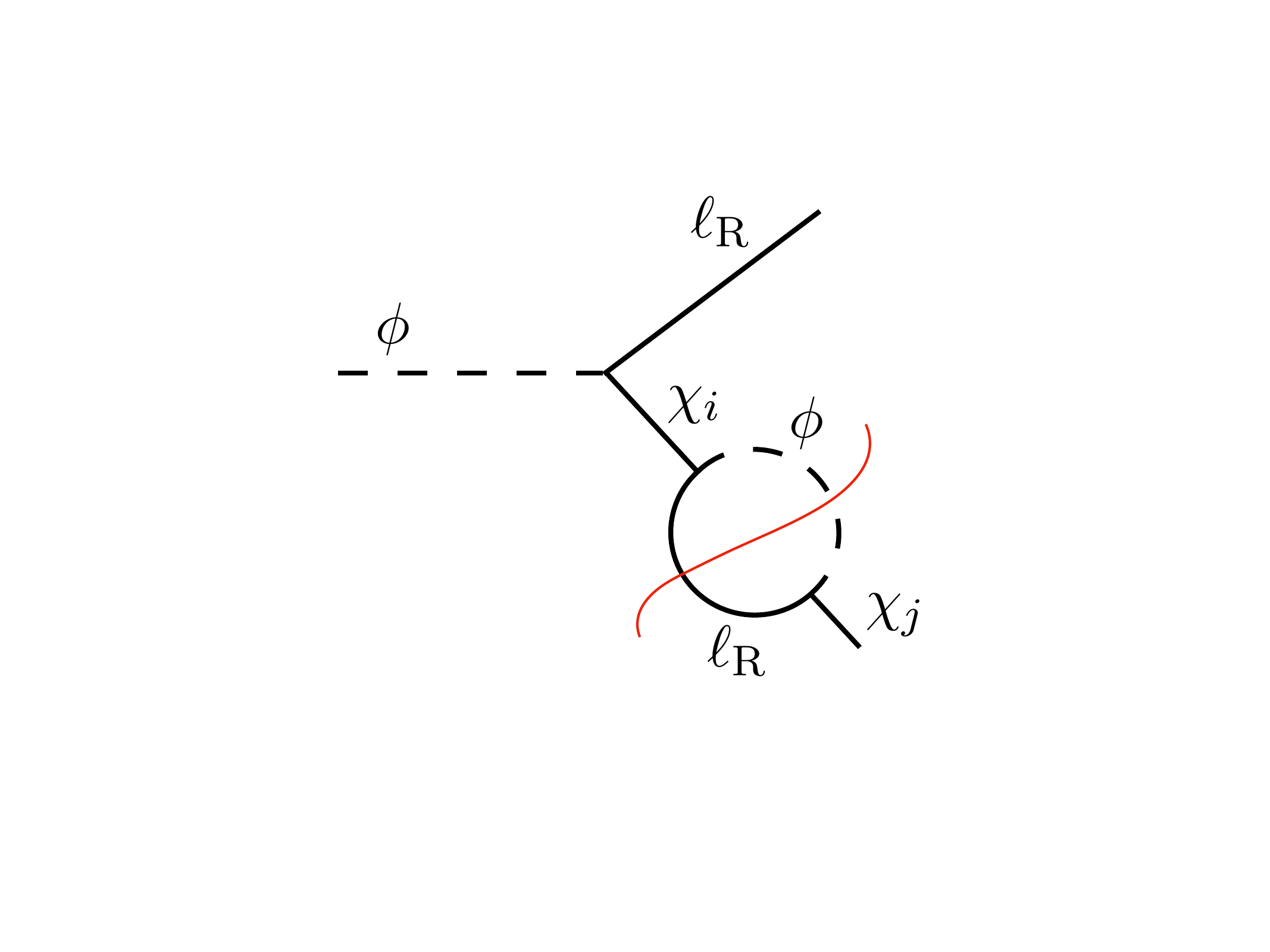}
    \vspace{-2ex}
    \caption{One loop self-energy diagram with thermal cut leading to the $CP$-asymmetry in the decay $\phi\to \chi_j \ell_\text{R}.$
    }
    \label{fig:asymdiag}
\end{figure}

The leading contribution to $\epsilon$ arises from the interference of the tree-level and one-loop diagram the latter of which is shown in Fig.~\ref{fig:asymdiag}. While the $CP$-asymmetry vanishes at zero temperature for the relevant parameter region $m_\phi+m_\ell>m_{\chi}$, the absorptive part is non-zero when accounting for thermal corrections. At finite temperatures, the $\phi$ or $\ell_\text{R}$ running in the loop can arise from the thermal bath, kinematically allowing the displayed cut~\cite{Giudice:2003jh,Garbrecht:2010sz,Frossard:2012pc,Hambye:2016sby}.
Given the small couplings $\sim 10^{-6}$, the required values of $\epsilon$ can only be achieved in the resonant regime of quasi-degenerate $\chi_i$. 
The respective process has been computed in the context of low-scale seesaw leptogenesis scenario for the analogous decay of a SM Higgs into an active and sterile neutrino, yielding~\cite{Hambye:2016sby}:
\begin{equation}
    \label{eq:epsHambye}
    \epsilon(T) = I_1 \frac{\xi \,\gamma(T)}{\left(\xi+\pi \varphi/(2 x)^2 \right)^2+\gamma(T)^2}\simeq I_1 \frac{ \gamma(T)}{\xi},
\end{equation}
where $I_1=\Im[(\lambda^\dagger\lambda)^2_{12}]/(|\lambda^\dagger\lambda|_{11}|\lambda^\dagger\lambda|_{22})$, $\xi=2 \Delta m_{12}/\Gamma^{0}_{\!\chi_2}$, 
and $\varphi$ is an $\mathcal{O}(1)$ number parametrizing thermal corrections to the mass splitting $\Delta m_{12}$. The product $\Gamma^{0}_{\!\chi_2} \gamma(T)$ corresponds to the absorptive part of the self-energy, where $\Gamma^{0}_{\!\chi_2}$ corresponds to the $\chi_2$'s would-be vacuum decay rate (in the limit $m_\phi,m_\ell\to0$) and $\gamma(T)$ is the spectral loop integral. Its integrand is proportional to $f_\ell+f_\phi$, \ie~the sum of the thermal distribution functions of the lepton and scalar of the cut propagators in Fig.~\ref{fig:asymdiag}, see Appendix D of Ref.~\cite{Frossard:2012pc} for details. Accordingly, $\gamma(T)$ is a steeply falling function in $x=m_\chi/T$, vanishing for $x\to\infty$. The approximation in Eq.~\eqref{eq:epsHambye} applies in the case $\gamma,\varphi\ll \xi$ which we find to be sufficiently justified in the semi-relativistic regime $x\gtrsim 2$ and for $\Gamma^{0}_{\!\chi_2}\ll \Delta m_{12}$. 
In this regime, the 
Boltzmann equations are expected to provide a good approximation. However, for $m_\chi \lesssim 200\,\text{GeV}$, its validity becomes questionable so we display results for the asymmetry only for $m_\chi> 200\,\text{GeV}$.
A more rigorous computation utilizing a density matrix formalism goes beyond the scope of this work. 
Note that in contrast to the seesaw scenario, the $\chi_i$ do not mix with the active neutrinos, remaining pure SM singlets. 

\begin{figure}
  \centering
\vspace{2ex}
\includegraphics[width=0.45\textwidth, trim= {0.0cm 0.0cm 0.0cm 0.06cm}, clip]{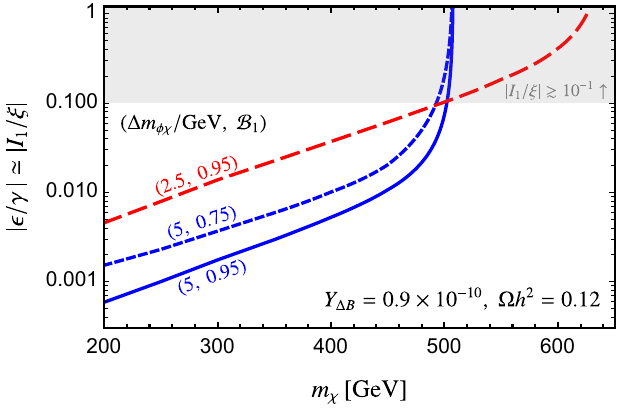} 
\vspace{-1ex}
  \caption{Required $|\epsilon/\gamma|\simeq |I_1/\xi|$ matching $Y_{\Delta B}^\text{meas}$ for three slices of parameter space with different $\Delta m_{\phi\chi}$ and $\mathcal{B}_1$ for 
$\lambda_H=1$. The overall coupling strength is adjusted such that $\Omega h^2=0.12$. The light gray area flags the regime of questionable validity (for concreteness we display $|\epsilon/\gamma|>0.1$). 
}
  \label{fig:YBsols}
\end{figure}

Since $\gamma(T_\text{Sphal})$ depends on $m_\chi$ and $m_\phi$ only, we can compute $\gamma \Delta_\phi/\epsilon$ at $T=T_\text{Sphal}$ for our solution of the Boltzmann equations and solve Eq.~\eqref{eq:YDeltaB} for the required value of $\epsilon/\gamma\simeq I_1/\xi$ that yields $Y_{\Delta B}=Y_{\Delta B}^\text{meas}$. Note that $I_1/\xi$ is a combination of temperature-independent model parameters only which \textit{a priori} can be as big as $\mathcal{O}(1)$. However, in our adopted regime of validity, $|I_1/\xi|\ll 1$.
Figure~\ref{fig:YBsols} shows 
this quantity as a function of $m_\chi$ for three viable parameter slices with different $\Delta m_{\phi\chi}$ and $\mathcal{B}_1$. 
The result demonstrates that conversion-driven leptogenesis can explain the DM density and BAU with realistic values  $|I_1/\xi|\sim(10^{-3}$--$10^{-1})$ in a significant part of the considered parameter space, up to $m_\chi\sim 500\,$GeV.

Note that for non-vanishing $I_1$, the mass splitting has to dominantly arise from a contribution beyond the minimal radiative (\ie~dark minimal flavor violation) one for which $\Delta m_\chi\propto \Re(\lambda^\dagger\lambda)$~\cite{Dev:2015wpa}. A quantitative evaluation of $\epsilon/\gamma$ including its dependence on the flavor structure of $\lambda$ is left for future work.


\vspace{1.6ex}
\emph{\bfseries Implications for searches}---Because of the small DM couplings required, $|\lambda_{ik}|\sim 10^{-6}$, the scenario is immune to canonical DM searches via direct and indirect detection. Similarly, the considered minimal model precludes measurable contributions to charge lepton flavor violation. For instance, ${\cal B}(\mu\to e \gamma)\propto \lambda^4$~\cite{Kersten:2014xaa} is more than ten orders of magnitude below the current limit~\cite{MEGII:2023ltw} and remains inaccessible in the foreseeable future~\cite{COMET:2018auw,Mu2e:2022ggl}.\footnote{Similarly, contributions to the lepton electric dipole moments are negligible, particularly as the one-loop contribution vanish due to the chiral coupling structure, see \eg~\cite{Cheung:2009fc}.}  
Three-flavor models may provide larger values though, if $\lambda$ is hierarchical.

At colliders, pairs of charged scalars can be produced at a sizable rate. Their decay into $\chi_i$ and $\ell_k$ leads to the signature of leptons plus missing energy (MET). At the LHC, corresponding searches have been performed in the regime of prompt decays in the context of supersymmetry, particularly in the compressed slepton-neutralino scenario~\cite{ATLAS:2019lng,CMS:2024gyw}. For non-prompt decays, displaced lepton (DL) searches~\cite{ATLAS:2020wjh,ATLAS:2023ios} and,  towards larger lifetimes, searches for disappearing tracks (DT)~\cite{ATLAS:2017oal,CMS:2020atg} and heavy stable charged particles (HSCPs)~\cite{CMS:2013czn,CMS-PAS-EXO-16-036,ATLAS:2019gqq} potentially constrain the model. However, none of these searches 
provide sensitivity to our scenario in the region $m_\phi> 140$\,GeV.\footnote{We use the production cross sections from Ref.~\cite{Bozzi:2007qr,Fiaschi:2019zgh} and utilize SModelS~\cite{Alguero:2021dig} for the reinterpretation of experimental results. 
}

\begin{figure}
  \centering
\vspace{1.5ex}
\includegraphics[width=0.45\textwidth]{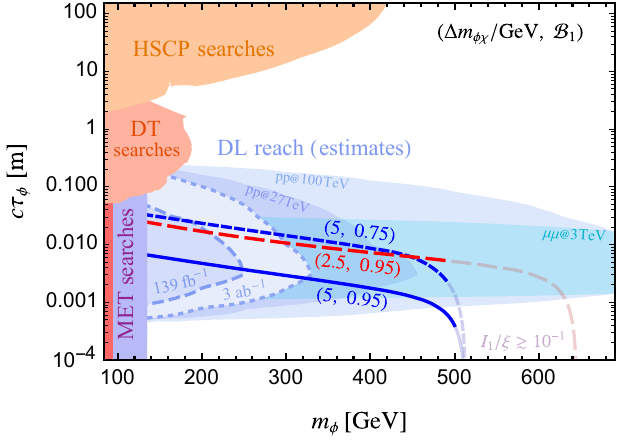}
\vspace{-1.5ex}
  \caption{
Decay length of the three cosmologically viable slices in parameter space (same as in Fig.~\ref{fig:YBsols}; the curves are shown fainted for $|\epsilon/\gamma| > 0.1$). The opaque shaded regions denote current 95\% CL exclusion limits from MET and HSCP searches at the LHC\@. The transparent regions show projections for the proposed soft DL searches, see main text.
}

  \label{fig:ctau}
\end{figure}

Figure~\ref{fig:ctau} shows the decay lengths of the three cosmologically viable slices in parameter space considered before. 
The royal blue and red shaded regions on the very left of the plot denote current limits from MET searches applying to mass splitting of 5 and 2.5\,GeV, respectively, assuming dominant decays into muons (limits for electrons are weaker). Towards large lifetimes stronger limits from DT and HSCP searches exist which are independent of decay products (orange shadings). 
In the regime of large asymmetries, $c\tau$ is typically of the order of a cm but ranges down to around a mm towards high masses. This region is most promisingly constrained by searches for DLs plus MET\@. Currently performed analyses, however, require relatively hard leptons, $p_\text{T}>65\,$GeV~\cite{ATLAS:2020wjh}, or di-leptons with $p_\text{T}>20\,$GeV and $m_\text{inv}>200\,$GeV~\cite{ATLAS:2023ios}. 

In our scenario, the displaced leptons are much softer -- typically of the order of the mass splitting, $\Delta m_{\phi\chi}$. Hence, the studied scenario can only be probed by 
lowering the $p_\text{T}$ (and $m_\text{inv}$) cut. A comparable level of SM background suppression may be achieved by exploiting the scenario's MET signature. Exploring specific MET-based discriminators or timing information of the tracker might further leverage background suppression in a high-luminosity environment in the future.
For illustration, we show the hypothetical reach of the DL search~\cite{ATLAS:2023ios} but with $p_\text{T}>10$\,GeV and no cut on $m_\text{inv}$ assuming that the same upper limits on the fiducial cross section can be achieved by a MET cut, $E_{\mathrm{T}}^{\mathrm{miss}}>200\,$GeV, as indicated by an estimate of the leading SM background of~\cite{ATLAS:2023ios}.
We show the 95\% CL exclusion reach for the current luminosity of the search, 139\,fb$^{-1}$, for a HL-LHC with 3\,ab$^{-1}$ and projections for a HE-LHC (pp@27TeV), FCC-hh (pp@100TeV) and future muon collider ($\mu\mu$@3TeV) for a similar integrated luminosity.\footnote{We use the cross section upper limits for the non-compressed scenario from Ref.~\cite{ATLAS:2023ios} and obtain the ones for the considered compressed scenario by applying the rescaling factor $\sigma_{\text{compr}}(p_\text{T}>10\,\text{GeV},E_{\mathrm{T}}^{\mathrm{miss}}>200\,\text{GeV})/\sigma_\text{non-compr}(p_\text{T}>10\,\text{GeV},m_{\mathrm{inv}}>200\,\text{GeV})$ from a simulation with MadGraph5\_aMC@NLO~\cite{Alwall:2014hca} and Pythia~8~\cite{Bierlich:2022pfr} employing merging~\cite{Lonnblad:2001iq} up to two jets. For the projection of the HL-LHC, we assume the background-uncertainty to be statistical. The future-collider projections are estimated by requiring the same production cross section as function of lifetime as the HL-LHC projection at the exclusion limit.
The limits assume $\Delta m_{\phi\chi}=5\,$GeV and decay into muons.}
These projections -- serving as rough estimates -- hint at the testability of the parameter space of interest, motivating dedicated phenomenological studies in the future.

While an indirect detection signal from DM annihilation is negligible, the emission or absorption of low-energy photons via the loop-induced process $\chi_2\leftrightarrow\gamma \chi_1$ constitutes an astrophysical prediction of the scenario. Interestingly, the mass splitting, $\Delta m_{12}\sim 0.01 m_\chi \lambda^2\xi $, is roughly in the meV range, providing a potential line signal in the infrared.
Note, however, that its rate is further suppressed by $\lambda^4$, severely challenging the observation of such a signal.


\vspace{1.6ex}

\emph{\bfseries Conclusions}---In the framework of conversion-driven freeze-out, the dark matter relic density is determined by semi-efficient conversion processes between dark matter and a pair-annihilating mediator. Despite their small rate, conversions are still large enough to thermalize DM at early times making the predictions robust against initial conditions. Remarkably, as demonstrated in this Letter, this scenario can concurrently produce the observed baryon asymmetry via \emph{conversion-driven leptogenesis}. 

We consider a minimal leptophilic dark matter model containing a Majorana dark matter flavor multiplet and a charged scalar mediator. 
The mild temperature dependence of the semi-efficient conversion rates causes a departure from thermal equilibrium of dark matter already at times when its abundance is still close to the relativistic one. This early out-of-equilibrium condition enables the generation of sizable lepton asymmetries produced by the same conversion processes that initiate freeze-out.  

Restricting ourselves to a minimal two-flavor model, we find viable parameter points, $\Omega h^2=0.12$, $Y_{\Delta B} \simeq 0.9\times 10^{-10}$, in the resonantly enhanced regime of quasi-degenerate dark matter multiplet states, reaching dark matter masses up to around 500\,GeV. Intriguingly, for a dark matter particle of a few hundred GeV, the evolution of the generated asymmetry develops a maximum close to the decoupling of sphaleron processes after which the converted baryon asymmetry is frozen. This parameter region is testable at colliders through the signature of displaced leptons with lifetimes in the millimeter to centimeter range. However, current searches that require comparably hard leptons lack sensitivity to our scenario, motivating dedicated searches for soft displaced leptons at the upcoming LHC runs and future colliders.

The parameter space is expected to open up further in the case of a three-flavor model. 
Moreover, introducing SU(2) couplings of the mediator resembles the scotogenic model which provides a natural explanation of small neutrino masses, in particular, in the conversion-driven scenario as recently shown in Ref.~\cite{Heeck:2022rep}. 
These findings merit future studies of different non-minimal realizations of the scenario exploring its links to the active neutrino sector. 


\medskip
\emph{\bfseries Acknowledgements}---I thank Marco Drewes, Mathias Garny, Thomas Hambye, Peter Maták, and Anil Thapa for valuable discussions and comments. I am particularly grateful to Julian Heeck for his insightful input, which helped me to improve the manuscript. I acknowledge support from the Alexander von Humboldt Foundation via the Feodor-Lynen Research Fellowship for Experienced Researchers and Feodor-Lynen Return Fellowship.


\bibliographystyle{bjstyle}
\bibliography{bib}

\end{document}